\documentstyle[12pt]{article}
\def\bold#1{\setbox0=\hbox{$#1$}%
      \kern-.025em\copy0\kern-\wd0
      \kern.05em\copy0\kern-\wd0
      \kern-.025em\raise.0433em\box0 }
\def\nsp{\noindent}
\def\nn{\nonumber}
\def\eea{\end{eqnarray}}
\def\bea{\begin{eqnarray}}
\def\eeas{\end{eqnarray*}}
\def\beas{\begin{eqnarray*}}
\def\ee{\end{equation}}
\def\be{\begin{equation}}
\def\bdm{\begin{displaymath}}
\def\edm{\end{displaymath}}
\def\fr{\frac}

\def\dag{^\dagger}

\def\fpi2{\mbox{F$_\pi$}^2}

\def\mpi2{{m_\pi}^2}
\def\mk{m_K}
\def\mk2{{m_K}^2}

\def\fk2{\mbox{F$_K$}^2}

\def\mss{{\eta}_{s,S}(q)}
\def\msk{{\eta}_{s,K}(q)}
\def\mvs{{\eta}_{v,S}(q)}
\def\mvk{{\eta}_{v,K}(q)}

\begin{document}
\begin{titlepage}
\begin{center}
\hfill FNT/T-95/16

\vspace*{2.0cm}
{\large\bf HYPERON RADIATIVE DECAYS IN THE BOUND STATE
SOLITON MODEL}
\vskip 1.5cm

{Carlos L. SCHAT$^a$, Carlo GOBBI$^b$,
and Norberto N. SCOCCOLA$^{a,c}$
\footnote[2]{Fellow of the CONICET, Argentina.} }
\vskip .2cm
{\it
$^a$ Physics Department, Comisi\'on Nacional de Energ\'{\i}a At\'omica,
          Av.Libertador 8250, (1429) Buenos Aires, Argentina. \\
$^b$ Department of Theoretical and
Nuclear Physics, University of Pavia, \\
and INFN, Sezione di Pavia, via Bassi 6, I-27100 Pavia, Italy.\\
$^c$ INFN, Sezione di Milano, via Celoria 16, I-20133 Milano, Italy.\\}
\vskip .2cm
\vskip 2.cm
May 1995
\vskip 2.cm
{\bf ABSTRACT}\\
\begin{quotation}
The radiative decays of hyperons are studied in the framework
of the bound state soliton model. Detailed predictions for the
total decay widths and the E2/M1 ratios corresponding to
decuplet--to--octet electromagnetic transitions are presented in
relation to future planned experiments at CEBAF and Fermilab. The
results are compared to those obtained in quark based models.
\end{quotation}

\end{center}
\end{titlepage}

In the near future, experiments at CEBAF\cite{Sch95} and Fermilab\cite{Rus95}
will provide new and precise information about the electromagnetic
decays of excited hyperons. Therefore, it is of great interest to
investigate these transitions in different theoretical frameworks
and compare various model predictions. In fact, transitions between
hadron states are more sensitive to their internal structure than hadron
spectra, and usually offer a more stringent test of hadron models than
``diagonal" properties like masses and magnetic moments.
Calculations of hyperon radiative decays have been available
for some time in the context of
the non-relativistic quark model \cite{DHK83,KMS85} and the MIT bag model
model \cite{KMS85}. More recently this type of transitions has also
been examined within heavy baryon chiral perturbation theory
\cite{BSS93} and a lattice quenched QCD simulation~\cite{LDW93}.
The purpose of the present article is to analyze the
hyperon radiative decays in the framework of the bound state soliton
model\cite{CK85,SNNR88}.  Here, we will concentrate on the
decuplet--to--octet electromagnetic transitions.
$\Lambda(1405)$ radiative decays as described in this model have been
studied in Ref.\cite{SSG95}. For the decuplet--to--octet decays
both M1 and E2 multipole transitions are allowed. Of course, unless they
are suppressed by some particular selection rule, the M1 transitions are
expected
to be strongly dominant over the E2 ones. Nevertheless, the ratio E2/M1 turns
out to be a rather important quantity since it reflects the possible existence
of charge deformations of the baryon states.
In fact, the E2/M1 ratio for the $\Delta \rightarrow N \gamma$ decay
has recently received considerable attention both theoretically (see e.g.
Refs.\cite{BSS93,LDW93,WW87} and references therein) and experimentally
(see e.g. Ref.\cite{DMW86}). For this reason, we will study
not only the total decay widths but also the E2/M1 ratios
corresponding to the hyperon radiative decays mentioned above.

In the bound state soliton model, one starts with
an effective $SU(3)$ chiral action which includes an appropriate
symmetry breaking term. We use
\be\label{lag}
\Gamma =\int d^4 x \Big\{- {f^2_\pi \over 4} Tr(L_\mu L^\mu) +
 {1\over{32 e^2}} Tr [L_\mu, L_\nu]^2 \Big\}
+ \Gamma_{\rm WZ} + \Gamma_{\rm sb}\, .
\ee
Here $\Gamma_{\rm WZ}$ is the non--local Wess-Zumino action
and $\Gamma_{\rm sb}$ is the
symmetry breaking term. Their explicit form can be found,
for instance, in Ref.~\cite{RS91}.  In Eq.(\ref{lag})
the left current $L_\mu$ is expressed in terms of the chiral field $U$
as $L_\mu=U^\dagger \partial_\mu U$.

Next, the Callan--Klebanov ansatz~\cite{CK85} is introduced
\be   \label{ansatz}
U=\sqrt{U_\pi} \, U_{K} \,  \sqrt{U_\pi} \ ,
\ee
where
\bea
U_K \ = \ \exp \left[ i\fr{\sqrt2}{{f_K}} \left( \begin{array}{cc}
                                                        0 & K \\
                                                        K\dag & 0
                                                   \end{array}
                                           \right) \right] \ , \
                                           \ \ \
K \ = \ \left( \begin{array}{c}
                   K^+ \\
                   K^0
                \end{array}
                           \right),
\eea
and $U_\pi$ is the soliton background field written as
a direct extension to $SU(3)$ of the $SU(2)$ field $u_\pi$, i.e.,
\bea
U_\pi \ = \ \left ( \begin{array}{cc}
                       u_\pi & 0 \\
                       0 & 1
                    \end{array}
                               \right ) \ ,
\eea
with $u_\pi$ being the conventional hedgehog solution
$ u_\pi=\exp[i \vec \tau \cdot \hat{ r}F (r) ]$.

According to the usual procedure, one expands up to the second order in the
kaon field. The Lagrangian density can therefore be rewritten as the sum
of a pure $SU(2)$ Lagrangian depending on the chiral field only and an
effective
Lagrangian describing the interaction between the soliton and the kaon
fields. The soliton profile is obtained by minimizing the classical
SU(2) energy while the kaon field satisfies the eigenvalue equation

\bea\label{eom}
\left[ -{1\over{r^2}} {d\over{dr}} (r^2 h {d\over{dr}} )
+ m_K^2 + V_{eff}^{\Lambda,l} -
f \ \omega_{\Lambda,l}^2 - 2 \ \lambda \ \omega_{\Lambda,l} \right]
k_{\Lambda,l}(r) = 0 \, ,
\label{eigen}
\eea

\noindent
where a mode decomposition of the kaon field in terms of the
grand spin $\vec \Lambda=\vec L + \vec T$ ($\vec L$ represents the
angular momentum operator and $\vec T$ is the isospin operator)
has been used.
In Eq.(\ref{eigen}) $\omega_{\Lambda,l}$ is the bound state energy for
given $(\Lambda,l)$.
The radial functions $h$, $f$, $\lambda$ and $V_{eff}^{\Lambda,l}$
are functions of  the chiral angle $F(r)$ only. Their explicit
forms can be found e.g. in Ref.\cite{SSG95}.

In this picture strange hyperons arise  as bound states of kaons to the
soliton. The octet and decuplet hyperons are obtained by populating
the lowest kaon bound state which carries the quantum numbers
 $\Lambda=1/2,\ l=1$.
The splittings among hyperons with different spin and/or isospin are given
by the rotational corrections, introduced according to the time--dependent
rotations:
\bea
u_\pi & \to & A u_\pi A^\dagger\, ,    \nonumber \\
K     & \to & A K \, . \label{rot}
\eea
This transformation adds an extra term of order $1/N_c$ to the Lagrangian.
The resulting mass formula which takes into account these rotational
corrections can be written as
\be
M_{I,J,{\cal S}} = M_{sol} + \omega \vert {\cal S} \vert +
{1\over{2\Theta}} \Big[ c J(J+1) + (1-c) I(I+1) +
{c(c-1)\over4}  \vert {\cal S} \vert (\vert {\cal S} \vert + 2 ) \Big] \, .
\ee
Here, $I$, $J$ and ${\cal S}$ are the isospin, spin and strangeness
hyperon quantum numbers respectively. $M_{sol}$ is the soliton mass,
$\Theta$ its moment of inertia and $c$ is the hyperfine splitting constant.
Their explicit form  are given, e.g. in Ref.~\cite{OMRS91}.

As mentioned above, in this work we are interested in the radiative
decays of the decuplet hyperons. Namely, in the processes
\bea
\Sigma^* &\rightarrow& \Lambda \ \gamma \nonumber \, ,\\
\Sigma^* &\rightarrow& \Sigma \ \gamma \nonumber  \, ,\\
\Xi^* &\rightarrow& \Xi \ \gamma \, .
\eea
For all these processes both M1 and E2 transitions are allowed.
Using the usual multipole expansion of the e.m. field~\cite{EG70},
the M1 partial decay width for these processes is given by
\bea
\Gamma_{M1} & = & 18 \ \alpha \ q\
 | < {\hat M}_3 (q)> |^2 \, ,
\label{m1}
\eea
\noindent
where the matrix element is taken between a decuplet hyperon
and an octet hyperon, both of them in states with spin projection
$J_3=+1/2$.
In Eq.(\ref{m1}), $\alpha =1/137$ is the e.m. fine structure constant
and $q$ is the photon momentum. The operator
${\hat M}_3 (q)$ is defined by
\bea
{\hat M}_3(q) &=& {1\over2}\  \epsilon_{3ij}\ \int d^3r \
{j_1(qr)\over{r}}\  r_i \ J^{em}_j \, ,
\eea
\noindent
where $j_1(qr)$ represents the $l=1$ spherical Bessel function
and $J^{em}_j $ are the spatial components of the electromagnetic
current.

On the other hand, the E2 partial decay width is given by
\bea
\Gamma_{E2} &=& {675\over8}\ \alpha \ q \ |<\hat Q_{33}(q)>|^2 \, ,
\label{e2}
\eea
\nsp
where the operator $\hat Q_{33}(q)$ is given by
\bea
\hat Q_{33}(q) &=& \int d^3r \
{j_2(qr)\over{r^2}} \ \left( z^2 - {r^2\over3} \right) \rho^{em} \, .
\eea
\nsp
Here $j_2$ represents the $l=2$ spherical Bessel function and
$\rho^{em}$ the electric charge density. It
should be noticed that in deriving Eq.(\ref{e2}) the Siegert's theorem
\cite{EG70}
has been used. For the typical photon momenta and
hyperon radii involved in the hyperon radiative decays we have
$qr \approx 1$. In this case, the condition
$j_{l-1}(qr) >> j_{l+1}(qr)$ necessary for the theorem to be valid,
is rather well satisfied for quadrupole transitions. One important
advantage of this method is that inconsistencies due to the
collective coordinate quantization of the soliton
are avoided\cite{WW87}.

In the bound state soliton model
the electromagnetic current $\vec J^{em}$ and charge
density $\rho^{em}$ are obtained from the
effective action by means of the Noether theorem. Given their explicit
forms the operators $\hat M_3(q)$ and $\hat Q_{33}(q)$ can be obtained
by using the Callan--Klebanov ansatz. After some algebra we get
\bea
{\hat M}_3(q) &=&  \mss \ J^S_3 + \msk \ J^K_3
         - 2 ( \mvs + |{\cal S}| \mvk ) \ R_{33}
\eea
\nsp
where $J^S$ the collective angular momentum, $J^K$ the bound kaon spin,
$R_{ij}={1\over2} Tr[\tau_i A \tau_j A^\dagger]$ and
\bea
\mss &=& - \frac{1}{3\pi {\Theta}} \int dr \ r j_1(qr)
          \sin^2 F F', \label{18} \\
\mvs &=& {2\pi\over3}  f_{\pi}^2 \int dr \  r j_1(qr)
             \sin^2 F \left[ 1 + {1\over{e^2 f_\pi^2}}
             \left( F'^2+ {\sin^2 F\over{r^2}} \right) \right], \label{21} \\
\msk &=& c \ \mss - \frac{2}{3}
 \int dr \  r j_1(qr) \left\{ k^2 \cos^2 \frac{F}{2} \right. \nonumber \\
& & \left. \hskip 0.5cm
+ \frac{1}{4 e^2 f_K^2} \left[ 4 \frac{k^2}{r^2} \sin^2 F
\cos^2 \frac{F}{2} + k^2 F'^2 \cos^2 \frac{F}{2} + 3 kk' F' \sin F \right]
\right\}, \label{19} \\
\mvk &=& \frac{1}{6} \int dr \ r j_1(qr) \left\{ k^2 \cos^2
\frac{F}{2} \left( 1 - 4 \sin^2 \frac{F}{2} \right) \right. \nonumber \\
& & \left. \hskip 0.5cm + \frac{1}{4 e^2 f_K^2}
 \left[ 4 \frac{k^2}{r^2} \sin^2 F \cos^2 \frac{F}{2}
\left( 3 - 8 \sin^2 \frac{F}{2} \right) \right. \right. \nonumber \\
& & \left. \hskip 1.5cm + k^2 F'^2 \cos^2 \frac{F}{2}
\left( 1 - 18 \sin^2 \frac{F}{2} \right) - 2 k^2 \omega^2 \sin^2 F \right.
\nonumber \\
& & \left. \left. \hskip 1.5cm
 + 2 k'^2 \sin^2 F + 3 kk' F' \sin F \left( 3 - 4 \sin^2 \frac{F}{2} \right)
\right] \right\} \nonumber \\
& & + \frac{N_c}{72} {\omega\over{f_K^2 \pi^2}}
\int dr \ r j_1(qr) \left( k^2 \sin^2 F F'
+ k k' \sin 2 F \right) \, . \label{22}
\eea
In the previous equations the subscripts $s$ and $v$ denote the isoscalar
and isovector parts, and $S$ and $K$ the pure soliton and kaon contributions,
respectively.

For $\hat Q_{33}(q)$ we obtain
\bea
\hat Q_{33}(q) &=& \nu_{v,S}(q)
                     \left[ J^S_3 \ R_{33} + {I_3\over3} \right]
                 + \nu_{v,K}(q) \left[ J^K_3 \ R_{33} -
                                  {J^K_a \ R_{3a}\over3} \right]
\label{q3}
\eea
\nsp
where $I_3$ is the z-component of the isospin operator and
$\nu_{v,S}(q)$ and $\nu_{v,K}(q)$ are
\bea
\nu_{v,S}(q) &=&  {8\pi f_\pi^2\over{15 \Theta}}
    \int dr \ r^2 \ j_2(qr) \ \sin^2F
    \left[ 1 + {1\over{e^2 f_\pi^2}} (F'^2 + {\sin^2F\over{r^2}} )
    \right] ,  \\
\nu_{v,K}(q) &=& c \ \nu_{v,S}  +
  {8\over{15}} \int dr \ r^2 \ j_2(qr)
    \left\{ \omega \ k^2 \cos^2{F\over2} \right. \nn \\
   & & \qquad \qquad  + \left. {\omega\over{4e^2 f_K^2}}
     \left[ k^2 \cos^2{F\over2} (F'^2 + 4 {\sin^2F\over{r^2}} )
            + 3 k k' F' \sin F \right] \right. \nn \\
 & & \qquad \qquad \left. - {N_c\over{12 \pi^2 f_K^2}}
        {\cos^2{F\over2}\over{r^2}} \left[ k^2 F' \cos^2{F\over2} -
            kk' \sin F \right] \right\} .
\eea

In order to calculate the corresponding matrix elements
of the transition operator $\hat M_3(q)$ and
$\hat Q_{33}$ we have to
evaluate the off-diagonal matrix elements of $J^S_3$, $J^K_3$, $I_3$
$R_{33}$, $J^S_3 R_{33}$, $J^K_3 R_{33}$ and $J^K_a R_{3a}$ between
hyperon wave functions. This is done using
standard angular momentum techniques. We obtain for the
matrix elements of $\hat M_3$
\bea
< \Lambda | \hat M_3 | \Sigma^*_0 > &=& {2\sqrt{2}\over3}
\left[ \mvs + \mvk \right] \, , \\
< \Sigma_0 | \hat M_3 | \Sigma^*_0  > &=& {\sqrt{2}\over3}
\left[ \mss - \msk \right] \, ,  \\
< \Sigma_\pm | \hat M_3 | \Sigma^*_\pm  > &=&
{\sqrt{2}\over3} \left[ \mss - \msk \pm
(\mvs + \mvk)  \right] \, , \\
< \Xi_{\mbox{{\footnotesize{\underline 0}}} }
 | \hat M_3 | \Xi^*_{\mbox{{\footnotesize{\underline 0}}} } > &=&
{\sqrt{2}\over3} \left[ \mss - \msk \pm {4\over3}
(\mvs + 2 \mvk)  \right]  \, ,
\eea
\nsp
while for the those of $\hat Q_{33}$
\bea
<\Lambda | \hat Q_{33}(q) | \Sigma_0^* > &=& - {\sqrt{2}\over6}
\nu_{v,S}(q) \, , \\
<\Sigma_0 | \hat Q_{33}(q) | \Sigma_0^* > &=& 0 \, , \\
<\Sigma_{\pm} | \hat Q_{33}(q) | \Sigma_{\pm}^* > &=& \mp
{\sqrt{2}\over6} \ \left[ \nu_{v,S}(q)
                 - { \nu_{v,K}(q)\over6 } \right] \, , \\
<\Xi_{\mbox{{\footnotesize{\underline 0}}}} |
\hat Q_{33}(q) | \Xi^*{\mbox{{\footnotesize{\underline 0}}}} >
&=& \mp  { 4\sqrt{2}\over{27} } \nu_{v,K}(q)  \, .
\eea

It should be noticed that when $qr << 1$ the Bessel
functions appearing in the expressions for the $\eta$'s and $\nu$'s
functions can be replaced by their small argument approximations.
In this case we have
\bea
\hat M_3(q) &\rightarrow& {q\over3} \ \mu_3 \, ,  \\
\hat Q_{33}(q) &\rightarrow& {q^2\over{15}} \ Q_{33} \, .
\eea
Here, $\mu_3$ is the static magnetic moment operator and
$Q_{33}$ the static electric quadrupole moment operator.
It should be kept in mind however that, as mentioned above,
$qr \approx 1$ for the transitions we are interested in.
In this case, therefore, this ``static" approximation is not
expected to be very good.

Finally we give the expressions for the E2/M1 ratios. This is
defined in terms of the matrix elements of the E2 and M1 transition
amplitudes as \cite{WW87}
\bea
{E2\over{M1}} &=& {1\over3} {<D(1/2)|M_{2,1}^{E2}|O(-1/2)>\over
                             <D(1/2)|M_{1,1}^{M1}|O(-1/2)>} \, .
\eea
\nsp
Here, $O(-1/2)$ represents an octet state with $J_3 = -1/2$ and
$D(1/2)$ a decuplet state with $J_3 = +1/2$. In terms of
the matrix elements of the $\hat M_3$ and $\hat Q_{33}$ operators
defined above this ratio can be expressed as
\bea
{E2\over{M1}} &=&  {5\over4} \ { <\hat Q_{33}>\over{<{\hat M_3}>}} \, .
\label{ratio}
\eea
Comparing this expression with Eqs.(\ref{m1},\ref{e2}) we note
that the following relation is satisfied
\bea
{\Gamma_{E2}\over{\Gamma_{M1}}} &=& 3 \left[ {E2\over{M1}} \right]^2\, .
\label{rela}
\eea
In the limit $qr<<1$ Eq.(\ref{ratio}) reduces to
the expression
\bea
{E2\over{M1}} &\rightarrow& {q\over4} \ { <Q_{33}>\over{<\mu_3>} }
\eea
given in Ref.\cite{WW87} for the particular case of the
$N \gamma \rightarrow \Delta$ process.

We turn now to the numerical calculations.
In order to estimate the uncertainties intrinsic in our
approach we will use two sets of values for the parameters appearing in the
effective action. In SET I we take $m_\pi=138 \ MeV$, $f_\pi=54 \ MeV$
and $e=4.84$. In SET II we consider massless pions and $f_\pi=64.5 \ MeV$
and $e=5.45$. In both cases we set the ratio $f_K/f_\pi$ and the kaon mass
to their empirical values $f_K/f_\pi = 1.23$ and $m_K = 495 \ MeV$.
Results for the total
decay widths are given in Table I. In all our calculations
we have taken the photon momentum $q$ as the empirical mass difference
between initial and final hyperon states\cite{DHK83,KMS85}.
We observe a good agreement between
our results and those obtained using the non-relativistic quark
model (NRQM) \cite{DHK83,KMS85,LDW93} and the bag model (BM)
\cite{KMS85} which are also listed in Table I. It
should be also noticed that the values
obtained by using heavy baryon chiral perturbation theory\cite{BSS93}
and quenched lattice QCD\cite{LDW93} span a range which is also
consistent with our predictions. This overall agreement between
different models contrasts with the situation for the $\Lambda(1405)$
decay widths\cite{SSG95} where the NRQM prediction is much larger
than the ones obtained with other models. This can be considered
as another indication that contrary to other low-lying hyperons the
$\Lambda(1405)$ can not be  simply understood as a 3-quark state.
Another interesting feature of our results is the strong suppression
of the $\Sigma_-^* \rightarrow \Sigma_- \gamma$ and
$\Xi_-^* \rightarrow \Xi_- \gamma$ in agreement with the
well-known $SU(3)$ $U$-spin selection rule\cite{Lip73}.
This might be understood by noting that, although in the bound
state approach strangeness degrees of freedom are treated
rather differently from the isospin ones, at the level of
the effective lagrangian the $SU(3)$ breaking is still not so
strong.
In Table I we have also included the decay width of the
only allowed octet--to--octet hyperon transition, namely
$\Sigma_0 \rightarrow \Lambda \gamma$ which is, of course, purely M1.
Presently, this is the only radiative hyperon decay for which
the empirical decay width is accurately known.
As we see, in this case
the soliton model prediction is reasonable good.
It is worthwhile to mention that, in the static limit, the corresponding
transition magnetic moment has already been evaluated in Ref.\cite{KM90}.

The predicted $E2/M1$ ratios for the decuplet--to--octet transitions
are given in table II. As expected, for the ``$U$-spin allowed''
transitions we obtain rather small values for the corresponding
ratios. On the other hand, for the ``$U$-spin forbidden''
processes $\Sigma_-^* \rightarrow \Sigma_- \gamma$ ,
$\Xi_-^* \rightarrow \Xi_- \gamma$ they are quite large.
Similar results\footnote{Our results have an overall
sign difference with respect to those quoted in
Refs.\cite{BSS93,LDW93}. This is probably due to a difference in the
way the E2/M1 ratio is defined.
Here, we follow the conventions used in Refs.\cite{WW87,EG70}.} have been
reported in
Refs.\cite{BSS93,LDW93} although our values for the forbidden
transitions are larger. Since for these transitions both the $M1$ and
$E2$ amplitudes are quite small, the precise values of the $E2/M1$
become more sensitive to the details of each model and quantitative
differences are expected to happen. A particular feature of our
predictions
is the vanishing of the E2 amplitude (and therefore of the
E2/M1 ratio) corresponding to the
$\Sigma_0^* \rightarrow \Sigma_0 \gamma$ decay. This is a consequence
of the spherical symmetry of the isoscalar charge density in the
soliton model. As it can be observed in Eq.(\ref{q3}) the operator
$\hat Q_{33}$ has only isovector components and therefore its
matrix elements between states with the same isospin turn out to
be proportional to the isospin projection.

In conclusion, we have studied the radiative transitions
between decuplet and octet hyperons within the bound
state soliton model. We have found that the predictions
for the full decay widths are in reasonable agreement with those
obtained in quark based models. In particular, the ``$U$-spin"
selection rule is quite well satisfied. Predictions for the $E2/M1$
ratios have also been made. Corresponding values for the
$\Sigma_-^* \rightarrow \Sigma_- \gamma$ and
$\Xi_-^* \rightarrow \Xi_- \gamma$ have been found to be quite
large. Experimental verification of these predictions as well as
those on the e.m. decays of the other excited hyperons will
certainly help to improve our understanding on the structure
of the hyperons.
\begin{center}
* \qquad * \qquad *
\end{center}
The authors wish to thank W. Weise for useful discussions
and G.L.Thomas for a
private communication on Eq.(\ref{22}). Most of the
work reported here was done while two of them (CLS and NNS) were
participating at the INT-95-1 Session on ``Chiral Dynamics in Hadrons
and Nuclei" at the Institute for Nuclear Theory at the University of
Washington, USA. They wish to thank the organizers of the Session
for the invitation to participate at it and the Department of Energy, USA
for partial financial support during that period.
CLS was partially supported by Fundaci\'on Antorchas and
CG by University of Pavia under
the Postdoctoral Fellowship Program.

\vspace*{1.cm}

\begin{center}
{\Large \bf Table I}
\vspace{0.5cm}

\begin{tabular}{|c|c|c|c|c|c|} \hline
    & \multicolumn{2}{|c|}{This work} & NRQM & BM &EMP\\ \cline{2-3}
    & SET I & SET II &   &    &   \\
 \hline
$\Sigma_0 \rightarrow \Lambda \gamma $ &
  8 &   7   & 8.5 & 4.6 &  $8.9 \pm 0.7$\\
$\Sigma^*_0 \rightarrow \Lambda\gamma $ &
243 & 170   & 273 & 152 &    \\
$\Sigma^*_0 \rightarrow \Sigma_0 \gamma $ &
 19 &  11   &  18 & 15  &   \\
$\Sigma^*_+ \rightarrow \Sigma_+ \gamma $ &
 91 &  59   & 110 &     &   \\
$\Sigma^*_- \rightarrow \Sigma_- \gamma$ &
  1 &   1   & 2.5 &     &   \\
$\Xi^*_0 \rightarrow \Xi_0 \gamma$ &
148 &  97   & 135 &     &  $   $  \\
$\Xi^*_- \rightarrow \Xi_-\gamma $ &
  5 &   5   & 3.2 &     &    \\ \hline
\end{tabular}
\vspace{0.5cm}

\begin{quotation} 
\noindent Hyperon radiative decay widths (in $keV$) as
calculated in the bound state soliton model. Note that the
partial M1 and E2 widths can be obtained from the total
widths listed here by using Eq.(\ref{rela})
together with the predictions for E2/M1 listed in Table II.
Also listed are the predictions of the non-relativistic quark
model (NRQM)\cite{DHK83,KMS85,LDW93} and bag model (BM) \cite{KMS85}.
The predicted $\Sigma_0 \rightarrow \Lambda \gamma$ decay
width is also included together with the corresponding empirical value.
\end{quotation}

\end{center}

\vspace*{2.cm}

\begin{center}
{\Large \bf Table II}
\vspace{0.5cm}

\begin{tabular}{|c|c|c|} \hline
    & SET I &  SET II  \\ \hline
%
%
$\Sigma^*_0 \rightarrow \Lambda$  & - 4.56  & - 5.43 \\
$\Sigma^*_0 \rightarrow \Sigma_0$ &  0     &  0    \\
$\Sigma^*_+ \rightarrow \Sigma_+$ & - 4.84  & - 7.61  \\
$\Sigma^*_- \rightarrow \Sigma_-$ & - 57.7  &  - 51.1 \\
$\Xi^*_0 \rightarrow \Xi_0$       & - 3.13  &  - 4.38 \\
$\Xi^*_- \rightarrow \Xi_-$       & - 17.8  &  - 18.5 \\
\hline
\end{tabular}
\vspace{0.5cm}

\begin{quotation}
\noindent Ratios E2/M1 (in \%) as calculated in the bound
state soliton model.
\end{quotation}

\end{center}

\end{document}